
\magnification=\magstep1
\baselineskip=20pt
\centerline{The Eddington Limit and Soft Gamma Repeaters}
\medskip
\centerline{J. I. Katz}
\centerline{Department of Physics and McDonnell Center for the Space
Sciences}
\centerline{Washington University, St. Louis, Mo. 63130}
\centerline{I: katz@wuphys.wustl.edu}
\bigskip
Observed intensities and inferred distances of soft gamma repeaters imply
luminosities in excess of the nominal (electron-scattering opacity)
Eddington limit by four to six orders of magnitude.  I review the physical
basis of this limit.  Accretional luminosities may exceed it if energy is
hydrodynamically coupled from accreting matter to closed field lines where it
forms a pair gas.  This magnetically confined pair gas radiates roughly a
black body spectrum with $k_B T_e \approx 23$ KeV, consistent with
observations, at a luminosity up to $\sim 3 \times 10^{44}$ erg s$^{-1}$ for
a surface field of $10^{13}$ gauss.  Magnetic transparency is not required.
I discuss the minutes-long continuing emission of March 5, 1979, steady
counterparts to SGR, their spin periods and the recoil problem.
\vfil
\noindent
Subject headings: gamma rays: bursts---stars: neutron---accretion
\eject
\centerline{1. Introduction}
\medskip
The luminosities of the three known soft gamma repeaters (SGR) during
outbursts are several orders of magnitude greater than the Eddington limit
of $\approx 2 \times 10^{38}$ erg s$^{-1}$ of a 1.4 $M_\odot$ neutron star
with the usual ($\approx 0.34$ cm$^2$ g$^{-1}$) electron scattering opacity
of stellar material.  For example, the highest luminosity observed from SGR
1806-20 was $1.8 \times 10^{42}$ erg s$^{-1}$ (Fenimore, Laros and Ulmer
1994), while SGR 0525-66 radiated about $5 \times 10^{44}$ erg s$^{-1}$
(Mazets, {\it et al.}~1979) at the peak of its outburst of March 5, 1979.
The purpose of this paper is to explain how SGR can produce such
extraordinarily high luminosities, particularly if their energy source is
accretion.  In \S2 I review the derivation and significance of the Eddington
limit, and discuss its applicability to and implications for three classes
of SGR models: accretional, thermonuclear and flare.  \S3 describes how
large violations of the Eddington limit may be possible in accretional
models.  In \S4 I consider the effects of magnetic fields: pair plasma may
be magnetically confined by a $10^{13}$ gauss surface field up to
luminosities $\sim 3 \times 10^{44}$ erg s$^{-1}$, but magnetic transparency
is not a satisfactory explanation of the super-Eddington luminosities of SGR.
Finally, \S5 contains a general discussion, including an explanation of the
extended emission observed on March 5, 1979 and the implications of the
recently discovered steady X-ray sources coincident with two SGR.
\bigskip
\centerline{2. The Eddington Limit}
\medskip
Eddington derived his famous limit as a consequence of the equations of
hydrostatic stellar structure (Chandrasekhar 1939).  The equation of
hydrostatic equilibrium is
$${dp(r) \over dr} = - {GM(r) \rho(r) \over r^2}, \eqno(1)$$
where $M(r)$ is the mass interior to the radius $r$, $\rho(r)$ is the
density and $p(r)$ is the pressure.  An elementary treatment of radiative
transfer at large optical depth yields a radiative luminosity
$$L = - {4 \pi r^2 c \over \kappa \rho} {dp_r \over dr}, \eqno(2)$$
where $\kappa$ is the Rosseland mean opacity and $p_r = (1 - \beta)p$ is the
radiation pressure.  Substituting (1) in (2) gives, for constant $\beta$,
the classical Eddington limit $L_E$:
$$L = {4 \pi c GM \over \kappa} (1 - \beta) < {4 \pi c GM \over \kappa}
\equiv L_E. \eqno(3)$$

This derivation begins with the equation of hydrostatic equilibrium; the
flow of radiation is secondary to the contribution of radiation pressure to
hydrostatic balance.  If $\kappa$ were to be increased instantaneously, the
consequence would be a reduction in the emergent luminosity, not the
expulsion of matter in a ``super-Eddington wind''.  Such a wind is
impossible because there is no regulatory mechanism to hold the radiative
luminosity constant in an optically thick star.  Inefficient convection, an
increase of opacity above a photosphere, or non-radiative heating of a
corona may drive a wind, but these processes are outside the scope of
Eddington's derivation because the assumptions of hydrostatic equilibrium
or of radiation diffusion fail; the radiative $L$ does not exceed $L_E$ in
optically thick regions.

The Eddington limit is also applicable to accretion flows.  Katz (1977,
1987) showed that in spherical accretion onto a body of mass $M$ and radius
$R$ the rate $v_{diff}$ of radiation diffusion outward through the accreting
matter at the surface satisfies
$${v_{diff} \over v_{ff}} = {{\dot M}_E \over 6{\dot M}}, \eqno(4)$$
where the free-fall velocity $v_{ff} = (2GM/R)^{1/2}$ and ${\dot M}_E
\equiv L_E R/(GM)$ is the accretion rate providing the luminosity $L_E$. A
super-Eddington luminosity ($L > L_E$) cannot emerge, even if ${\dot M} >
{\dot M}_E$, because the radiation is trapped by the opaque accreting matter
within which it is produced.  This conclusion was verified in spherically
symmetric (Klein, Stockman and Chevalier 1980; Burger and Katz 1980, 1983)
and disk (Eggum, Coroniti and Katz 1988) accretion calculations.

The Eddington limit is inapplicable to explosions, which are not in
hydrostatic equilibrium, and which become transparent because their flows
diverge spherically.  As a result, supernovae ($L \sim 10^{44}$ erg
s$^{-1}$) and classical gamma ray bursts (GRB) produced by fireballs at
cosmological distances ($L \sim 10^{51}$ erg s$^{-1}$, the most
electromagnetically luminous objects in the universe) may exceed their
Eddington limits by many orders of magnitude.  In contrast, the fact that
novae have luminosities comparable to $L_E$ is evidence that they are
produced by nearly hydrostatic configurations, rather than by explosions;
their mass outflow is a consequence of disruption of the erupting star's
envelope by its binary companion, and is incidental rather than essential;
if the envelope is not disrupted (because the orbit is large) a nova-like
variable or symbiotic star is observed.

No Eddington limit is applicable to energy transported non-radiatively, such
as by convection in stellar interiors.  It is for this reason that the
envelopes of cool stars are convective; their opacity is very large and the
local $L_E$ (defined by the local opacity) is less than the stellar luminosity.

An Eddington limit (3), with $\kappa \ge \kappa_{es} \approx 0.34$ cm$^2$
g$^{-1}$ and $L_E \le 2 \times 10^{38}$ erg s$^{-1}$, appears to constrain
thermonuclear and accretional models of SGR.  The characteristic dynamical
time of a neutron star ($< 10^{-4}$ s) is much shorter than the observed
durations of SGR ($\sim 0.1$ s), implying hydrostatic equilibrium (1).
Large optical depths (in thermonuclear models because of the required
temperature and density, and in accretional models because their luminosity
requires ${\dot M} \gg {\dot M}_E$) imply the diffusion of radiation (2).

The Eddington limit is not applicable to flare models, because in them energy
transport is magnetohydrodynamic rather than by diffusion of radiation.  The
radiating plasma must be confined against radiation pressure, but if it is
optically thin confinement can be provided by a neutron star's magnetic
field (Katz 1982, 1993, 1994), without the plasma trapping the radiation.
Unfortunately, the present understanding of flares is inadequate to make
specific predictions and flare models are nearly infinitely flexible; in
addition, observed Solar flares resemble less well the single-peaked
profiles and thermal spectra of SGR than the more complex structures and
nonthermal spectra of classical GRB.
\goodbreak
\bigskip
\goodbreak
\centerline{3. Pair Gas and Neutron Star Accretion}
\medskip
In order for an accretion flow to radiate in excess of $L_E$ the
accretional luminosity must be extracted from the accreted matter itself by
a process other than diffusion of radiation.  I suggest that this may be
accomplished magnetohydrodynamically.  Suppose a discrete body of accreting
matter (such as the planetary fragments considered by Colgate and Petschek
1981 and Katz, Toole and Unruh 1994) falls onto a magnetic neutron star.  At
impact the accreted matter has speed $v_{ff} \approx c/2$ and density $\rho
\sim$ 8--100 g cm$^{-3}$, and supplies a dynamic stress $\rho v_{ff}^2 \sim
10^{21}$--$10^{22}$ dyne cm$^{-2}$.  This stress is $\sim 3$--30 times less
than the magnetic stress of a $10^{12}$ gauss field, so that if the neutron
star has a magnetic moment $\sim 10^{30}$--$10^{31}$ gauss cm$^3$ (as is
typical for pulsars) the flow within a few radii of its surface is strongly
affected by the field.

The characteristic wave speed in a vacuum magnetosphere, or in one filled
with pair gas, is $\sim c$.  Because this is only about twice the infall
speed, the coupling efficiency of infall kinetic energy to magnetospheric
disturbances may be as large as $\epsilon_{SGR} \sim v/c \sim 0.5$,
as would be estimated for an elastic collision between the infalling fluid
and a magnetohydrodynamic wave.  Energy coupled to the magnetosphere
produces a varying magnetic field, and hence an electric field.  The
amplitude of this electric field may approach that of the magnetic field
(because the hydrodynamic stress approaches the magnetic stress, and the
characteristic power density approaches $B^2 c/(8 \pi)$), which is large
enough to produce a vacuum pair breakdown cascade (Epstein and Smith 1993).
The detailed electrodynamics is complex and beyond the scope of this paper;
I simply assume that a fraction $\epsilon_{SGR}$ of the accretional power
appears as pair plasma on the closed magnetic field lines, where it can be
contained by the magnetic stress.

The Thomson scattering opacity of pair plasma (referred to the pair rest
mass only)
$$\kappa_\pm = {\sigma_{es} \over m_e} = {8 \pi \over 3}{e^4 \over m_e^3
c^4} = 726\ {\rm cm}^2\ {\rm g}^{-1}, \eqno(5)$$
where $\sigma_{es}$ is the Thomson scattering cross-section, is extremely
high, and the corresponding Eddington limit for a $1.4\ M_\odot$ neutron
star would be only $1.0 \times 10^{35}$ erg s$^{-1}$.  However, the
particle density of a nondegenerate equilibrium pair plasma with zero net
lepton number is a sensitive function of temperature:
$$n_\pm = {2 (2 \pi m_e k_B T)^{3/2} \over h^3} \exp(-m_e c^2/k_B T).
\eqno(6)$$
The volumetric opacity (beam attenuation per unit length) is $2 \sigma_{es}
n_\pm$, where the factor of two comes from the presence of both e$^+$ and
e$^-$; this is a steeply increasing function of temperature for $k_B T \ll
m_e c^2$.

At high temperatures the particle density of an equilibrium pair gas (6) and
its volumetric opacity are high, and if it is confined the emergent
luminosity will be small.  At lower temperatures $n_\pm \to 0$, the plasma
is transparent, and its energy density (almost entirely radiative) escapes
freely even if the charged particles are magnetically confined.  A
photosphere is defined by the usual condition on the optical depth
$$\tau = 2 \sigma_{es} n_\pm \lambda = {2 \over 3}, \eqno(7)$$
where $\lambda$ is a characteristic density scale length.  Adopting $\lambda
= 10^5$ cm in (7) and using (6) yields a characteristic temperature
$$T_c \approx 2.7 \times 10^{8\ \circ}{\rm K}. \eqno(8)$$
This value of $T_c$ is almost independent of $\lambda$; $d\ln T_c/d\ln
\lambda \approx 0.042$.

A nonlinear cooling wave (analogous to an inverse Marshak [1958] heating
wave) will penetrate a volume of magnetically confined pair plasma, with its
visible photosphere at an effective temperature $T_e \approx T_c$.  The
velocity of this cooling wave is given by energy balance arguments as
$$v_{cool}(T) = {\sigma_{SB} T_e^4 \over a T^4} < {c \over 4}, \eqno(9)$$
where $a$ is the radiation constant and $\sigma_{SB}$ the Stefan-Boltzmann
constant.  Regions hotter than $T_c$ are opaque and have negligible energy
flow; they are hidden from direct observation by their own opacity.  Cooler
regions are transparent; their $n_\pm$ and emissivity are negligible, and
they contain chiefly free-streaming radiation.

The emergent spectrum will be {\it roughly} that of a black body at the
temperature (8).  This spectrum is similar to that typically seen in SGR,
and explains the observation of Fenimore, Laros and Ulmer (1994) that the
spectral shape of outbursts of SGR 1806-20 is independent of intensity, with
only the radiating area varying between outbursts.  Quantitative predictions
of the spectrum require a detailed radiative transfer calculation, including
the effects of recoil in Compton scattering and cyclotron and cyclotron
harmonic opacity and polarization, and will depend on the magnitude and
direction of the magnetic field.  The observed peak spectral frequencies of
SGR are comparable to the cyclotron fundamental or low harmonic frequencies
for $10^{12}$--$10^{13}$ gauss fields, and substantial elliptic polarization
is likely.

A characteristic luminosity may be defined:
$$L_c \equiv 4 \pi R^2 \sigma_{SB} T_c^4 \approx 4 \times 10^{42}\ {\rm erg}
\,{\rm s}^{-1} \gg L_E. \eqno(10)$$
SGR need not have luminosities this large; those less luminous may radiate
from only a fraction of the neutron star's surface, while those with greater
luminosity may have an effective pair gas photosphere with greater area than
the neutron star's surface.  The
super-Eddington luminosity (10) is permitted by the transparency of vacuum;
the important effect of the magnetic field is to provide a confining stress,
preventing relativistic expansion.

The harder spectrum observed (Mazets, {\it et al.}~1979) during the initial
peak of the outburst of March 5, 1979 may be explained if during that brief
($\sim 0.1$ s) period the pair plasma had too great an energy density to be
magnetically confined, and expanded relativistically; the transparency
wave occurs at the same co-moving temperature $T_c$ (8), but the Doppler
shift hardens the observed spectrum (Goodman 1986, Paczy\'nski 1986).
However, an essential property of the model presented here is that $L \gg
L_E$ does not require an expanding fireball, and only in the first $\sim
0.1$ s of the burst of March 5, 1979 was such a fireball likely to have been
observed.
\bigskip
\goodbreak
\centerline{4. Magnetic Fields}
\medskip
The magnetic field of SGR 0525-66 may be constrained by the observation of a
period $P = 8$ s.  If this is interpreted as neutron star rotation, then the
radiation rate $2 \mu^2 \Omega_{NS}^4 / (3c^3)$ of a rotating dipole $\mu$
(and a similar expression for the wind of the aligned component of the dipole
moment) lead to the bound on the polar field:
$$B_p < \left({3 I c^3 \over \Omega_{NS}^2 R^6 t}\right)^{1/2} = 6.5 \times
10^{14} \left({I_{45} \over R_6^6 t_4}\right)^{1/2} P_8\ {\rm gauss},
\eqno(11)$$
where $t$ is the neutron star's age, $I$ is its moment of inertia,
$\Omega_{NS}$ is its spin angular velocity, $t_4 \equiv t/(10^4\,{\rm y})$,
$I_{45} \equiv I/(10^{45}\,{\rm g\,cm}^2)$, $R_6 \equiv R/(10^6\,{\rm cm})$,
and $P_8 \equiv P/(8\,{\rm s})$.

This argument could be applied to the other SGR if spin periods were
observed.  For example, if the suggested (Ulmer, {\it et al.}~1993) 2.8 s
period of SGR 1806-20 is confirmed then (11) would imply $B_p < 2.3 \times
10^{14} (I_{45}/R_6^6 t_4)^{1/2}$ gauss.
\medskip
\centerline{4.1 Magnetic transparency}
\smallskip
Canuto, Lodenquai and Ruderman (1971) calculated the Thomson scattering
opacity of strongly magnetized plasma with electronic gyrofrequency
$\Omega_e = e B/(m_e c)$ in excess of the photon frequency $\omega$, and
found that for the extraordinary mode the opacity is reduced from the
unmagnetized value by a factor $\sim (\omega/\Omega_e)^2$.  A similar
reduction factor will apply to other electronic contributions to the
opacity.  The Rosseland mean opacity in the presence of a strong magnetic
field is therefore $\sim 2 (\omega/\Omega_e)^2$ times the zero-field value,
where the factor of two results from the fact that the magnetic field
reduces the opacity of only one of the two polarization states.

Paczy\'nski (1992) and Duncan and Thompson (1994) pointed out that for
sufficiently large magnetic fields the observed luminosities of SGR may be
sub-Eddington, using the magnetically reduced opacity.  The required fields
are given by
$$B > {\omega m_e c \over e} \left({2 L \over L_0}\right)^{1/2}, \eqno(12)$$
where $L_0 \approx 2 \times 10^{38}$ erg s$^{-1}$ is the zero-field
Eddington limit.  For photons with $\hbar \omega = 70$ KeV (near the peak of
$\nu F_\nu$ for SGR) a luminosity $L \approx 1.8 \times 10^{42}$ erg s$^{-1}$
(observed for SGR 1806-20; Fenimore, Laros and Ulmer 1994) would require $B
> 8 \times 10^{14}$ gauss, while the peak luminosity of $5 \times 10^{44}$
erg s$^{-1}$ observed from SGR 0525-66 on March 5, 1979 (Mazets, {\it et
al.}~1979) would require $B > 1.3 \times 10^{16}$ gauss and its minutes-long
continued emission with initial $L \approx 3.6 \times 10^{44}$ erg s$^{-1}$
would require $B > 1.1 \times 10^{15}$ gauss.  These values are all
inconsistent with the bound (11), given the suggested spin periods.

For very large magnetic suppression of electron scattering opacity it is
necessary also to consider Thomson scattering by ions.  The ionic
cross-section
$$\sigma_{is} = {Z^4 \over A^2} \left({m_e \over m_p}\right)^2 \sigma_{es}.
\eqno(13)$$
The corresponding opacity for pure hydrogen $\kappa_{Hs} = 1.16 \times
10^{-7}$ cm$^2$ g$^{-1}$, while for pure iron $\kappa_{Fes} = 3.07 \times
10^{-7}$ cm$^2$ g$^{-1}$.  Assuming planetary cores to be iron, the
corresponding Eddington limit for a $1.4\ M_\odot$ neutron star, considering
ionic Thomson scattering opacity alone (supposing complete magnetic
suppression of electron scattering opacity), is $2.3 \times 10^{44}$ erg
s$^{-1}$, approximately two times too small to permit the observed peak
luminosity of SGR 0525-66 on March 5, 1979.  The ionic opacity may itself be
magnetically suppressed if $\Omega_i > \omega$, but for 70 KeV photons this
suppression begins around $B \approx 2.4 \times 10^{16}$ gauss, and $B > 4
\times 10^{16}$ gauss would be required to explain the observed $L \approx 5
\times 10^{44}$ erg s$^{-1}$.
\medskip
\centerline{4.2 Magnetic confinement}
\smallskip
The minimum magnetic field required to confine a source of size $\sim R$
radiating a luminosity $L$, continually resupplied, is given by comparison
of the radiative and magnetic stresses as
$$B > \left({2L \over R^2 c}\right)^{1/2}. \eqno(14)$$
If the radiation continues over a time $\tau$ and the radiated energy
$\sim L \tau$ was not resupplied but was all initially contained within a
region of volume $\sim 4 \pi R^3/3$ then the required field would be
$$B > \left({6 L \tau \over R^3}\right)^{1/2}. \eqno(15)$$
For $\tau \gg R/c$ the condition (14) is less demanding than (15) by a
factor which is typically $\sim 100$ for SGR.  The approximate consistency
between the observed duration $\sim 0.1$ s of SGR outbursts and the
calculated duration of infall of planetary fragments suggests that energy is
radiated at the rate at which it is supplied by accretion, so that (14) is
applicable.

Using in (14) black body radiation at an effective temperature equal to the
characteristic temperature (8) yields a unique value for the
required confining field $B_c$:
$$B > B_c \equiv \left({8 \pi \sigma_{SB} T_c^4 \over c}\right)^{1/2}
\approx 1.6 \times 10^{10}\ {\rm gauss}, \eqno(16)$$
a condition easily satisfied by most empirically inferred fields of neutron
stars.

The radiating photosphere may be larger than the neutron star's surface.
In this case, assuming a dipole for the confining field $B$, the condition
$B > B_c$ yields a maximum luminosity of confined pair plasma which
substantially exceeds that given by (10):
$$L_{max} \approx L_c \left({B_e \over B_c}\right)^{2/3} \approx 6 \times
10^{43} \left({B_e \over 10^{12}\ {\rm gauss}}\right)^{2/3}\
{\rm erg\ s}^{-1}, \eqno(17)$$
where $B_e$ is the surface equatorial field.  The increase of $L_{max}$ over
$L_c$ is entirely the consequence of the increased radiating area; $T_c$ is
unchanged.  The peak intensity of SGR 0525-66 on March 5, 1979 was too large
for magnetic confinement unless $B_e > 3 \times 10^{13}$ gauss, as suggested
by the interpretation (\S3) of its harder spectrum as the result of
relativistic expansion.  All other observed SGR intensities are consistent,
by ample margins, with magnetic confinement by neutron stars with typical
observed pulsar fields.
\bigskip
\centerline{5. Discussion}
\medskip
The hypothesis of magnetohydrodynamic coupling of accretional energy to a
magnetically confined pair plasma can solve the problem of the enormously
super-Eddington luminosities of SGR.  This solution is consistent with their
self-absorbed thermal spectra, whose temperatures are comparable to that
given by (8) and are nearly independent of luminosity.  Magnetohydrodynamic
coupling occurs naturally in accretional models, which may be able to
explain many of the other properties of SGR.

At the temperature (8) the thermal velocity of the barycenter of an
e$^+$-e$^-$ pair is about $0.15 c$.  This suggests the possibility of an
observable annihilation line, such as that reported by Mazets, {\it et
al.}~(1979).  However, they reported it only during the early part of the
outburst; the relativistic expansion which may have occurred at that time
would be inconsistent with an observable line.  Calculations by Carrigan and
Katz (1992) showed only a broad bump in the spectrum around 300 KeV, rather
than a recognizable annihilation line, because in their calculations photons
and pairs did not come to thermodynamic equilibrium, and the pair
temperature was far in excess of that given by (8).  They also found a
nonthermal high energy spectral tail because their assumed source was a
power law distribution of photons which extended throughout their grid,
including transparent regions far from the neutron star.

The 0.1 s intensity peak of March 5, 1979 was followed by a few minutes of
continued emission with slowly fading luminosity $\sim 10^{42}$ erg s$^{-1}$.
This may have been the result of continuing accretion, which probably
occurred if the distance of closest approach $R_c$ of the initial orbit of
the accreted fragment to the center of the neutron star lay between $R$ and
$\sim 5R$.  For a one-dimensional distribution of specific angular momentum
$\ell$ (as expected; Katz, Toole and Unruh 1994) the distribution of matter
in $R_c$ is $dM/dR_c \propto R_c^{-1/2}\,dM/d\ell \propto R_c^{-1/2}$, while
for a two-dimensional distribution $dM/dR_c$ is independent of $R_c$ and
for a three-dimensional distribution $dM/dR_c \propto R_c^{1/2}$.  In each
case the probability of a near-miss ($R < R_c < 5 R$) is at least comparable
to that of a direct hit ($R_c < R$).

The interaction between an accreting disrupted planetary fragment and the
neutron star's magnetic field is uncertain and complex.  In a near miss some
matter gives much of its angular momentum to the field and is steeply
channeled onto the stellar surface, producing the initial spike in
intensity.  Other matter loses only a little of its kinetic energy and
angular momentum, and re-emerges from the magnetosphere.  Its orbits remain
nearly parabolic, but their periods are shortened; returning matter arrives
in a continuous rain, and produces a continuous accretional luminosity.

If a fraction $\delta \ll 1$ of the specific angular momentum of an initially
parabolic orbit is lost by a tangential impulse, the semi-major axis of the
new orbit is $\approx R_c/(4\delta)$.  The new orbital period is $6 \times
10^{-5} (R_c / (\delta R))^{3/2}$ s; a return time of $t_r$ corresponds to
$\delta = 7 \times 10^{-5} (R_c/R) (t_r/100\,{\rm s})^{-2/3}$.  Because near
misses are expected to be about as frequent as direct hits, many
repetitions of SGR are predicted to be followed by emission of lower
intensity but longer duration and comparable fluence (Mazets, {\it et
al.}~1979 estimated that in the burst of March 5, 1979 approximately 25\% of
the total energy was radiated in the initial $\sim 0.1$ s, and 75\% in the
subsequent $\sim 100$ s).

The efficiency of conversion of accretional energy to pair plasma is
certainly less than unity, and very likely less than 0.5.  The majority of
the accretional energy is trapped in optically thick thermal plasma, and
will emerge at nearly the Eddington limit.  Recent observations have found
point X-ray source counterparts to SGR 0525-66 (Rothschild, Kulkarni and
Lingenfelter 1994) and SGR 1806-20 (Murakami, {\it et al.}~1994; Sonobe,
{\it et al.}~1994) with steady luminosities $\approx 10^{36}$ erg s$^{-1}$
and $\approx 3 \times 10^{35}$ erg s$^{-1}$ respectively.  These steady
luminosities exceed the expected luminosities of cooling neutron stars of
the estimated ages, and it is possible that they are associated with the
accretional energy released in the outbursts.  However, they are well below
the nominal (electron scattering) Eddington limit, and thus apparently
inconsistent with their interpretation as the gradual release of trapped
accretional energy.

The steady luminosity of SGR 1806-20 exceeds its time-averaged burst
luminosity $\sim 100$-fold.  The steady luminosity of SGR 0525-66 exceeds
its time-averaged burst luminosity (excluding March 5, 1979) by an even
larger factor but is comparable to the energy released on March 5, 1979
divided by the time since that date.  These facts also suggest that the
trapped accretional energy may be insufficient to power the steady
counterparts.

The classical power source of neutron stars is their rotational energy.  If
the apparent 8 second period of SGR 0525-66 is its spin period, then its
spin-down power $I \Omega_{NS}^2 / (2t) \sim 10^{33} / t_4$ erg s$^{-1}$,
inadequate to explain its observed $\sim 10^{36}$ erg s$^{-1}$ steady
counterpart.  A spin-down power supply would be adequate if $P < 0.3
t_4^{-1/2}$ s, or if the SNR contains a second neutron star with the
required smaller $P$.  The steady X-ray counterpart to SGR 1806-20
of $\sim 3 \times 10^{35}$ erg s$^{-1}$ similarly would require $P < 0.5
t_4^{-1/2}$ s, consistent with all known data (except the very uncertain
suggestion of a 2.8 s period by Ulmer, {\it et al.}~1993).

The requirement that the present spin-down power be supplied by a neutron
star of age $t$ may be used to constrain its magnetic field.  The neutron
star's spin rate is eliminated from (11) by requiring that the observed
X-ray luminosity $L_X \equiv 10^{36} L_{36}$ erg s$^{-1}$ be powered by
spin-down with efficiency $\epsilon_X \equiv 0.01 \epsilon_{0.01}$.
This yields a bound on the magnetic dipole moment
$$\mu < 1.0 \times 10^{30} {I_{45} \over t_4} \left({\epsilon_{0.01} \over
L_{36}}\right)^{1/2}. \eqno(18)$$

The presence of neutron stars with comparatively fast spin ($P < 0.3$ s)
would also explain the observation that the SNR are ``plerionic''
(center-filled) and apparently powered by a neutron star rather than by the
collision of stellar debris with interstellar gas.  The methods of
estimation of age applicable to collision-powered SNR (Kulkarni, {\it et
al.}~1994; Rothschild, Kulkarni and Lingenfelter 1994) would then be
inapplicable, and their actual ages could be significantly longer.  If $t >
10^5$ y then the inferred neutron star recoil velocities would be consistent
with the retention of planets, as required in models of SGR which assume
accretion of planetary fragments, even if the SGR are dynamically
off-center in their SNR.

I thank NASA NAGW-2918 for support and N.~G.~H.~Sh.~for illuminating these
apparently miraculous events.
\vfill
\eject
\centerline{References}
\parindent=0pt
\def\ref{\medskip \hangindent=20pt \hangafter=1}
\ref
Burger, H. L. and Katz, J. I. 1980, Ap. J. 236, 921
\ref
Burger, H. L. and Katz, J. I. 1983, Ap. J. 265, 393
\ref
Canuto, V., Lodenquai, J. and Ruderman, M. 1971, PRD 3, 2303
\ref
Carrigan, B. J. and Katz, J. I. 1992, Ap. J. 399, 100
\ref
Chandrasekhar, S. 1939, Introduction to the Study of Stellar Structure
(Chicago: U. Chicago Press)
\ref
Colgate, S. A. and Petschek, A. G. 1981, Ap. J. 248, 771
\ref
Duncan, R. C. and Thompson, C. 1994, in Gamma-Ray Bursts, ed. G. J. Fishman,
J. J. Brainerd and K. Hurley (New York: AIP), 625
\ref
Eggum, G. E., Coroniti, F. V. and Katz, J. I. 1988, Ap. J. 330, 142
\ref
Fenimore, E. E., Laros, J. G. and Ulmer, A. 1994, Ap. J. 432, 742
\ref
Goodman, J. 1986, Ap. J. 308, L47
\ref
Katz, J. I. 1977, Ap. J. 215, 265
\ref
Katz, J. I. 1982, Ap. J. 260, 371
\ref
Katz, J. I. 1987, High Energy Astrophysics (Menlo Park, Cal.:
Addison-Wesley)
\ref
Katz, J. I. 1993, in Compton Gamma-Ray Observatory, ed. M. Friedlander, N.
Gehrels, D. J. Macomb (New York: AIP), 1090
\ref
Katz, J. I. 1994, Ap. J. 422, 248
\ref
Katz, J. I., Toole, H. A. and Unruh, S. H. 1994, Ap. J. in press (December
20)
\ref
Klein, R. I., Stockman, H. S. and Chevalier, R. A. 1980, Ap. J. 237, 912
\ref
Kulkarni, S. R. {\it et al.} 1994, Nature 368, 129
\ref
Marshak, R. E. 1958, Phys. Fl. 1, 24
\ref
Mazets, E. P. {\it et al.} 1979, Nature 282, 587
\ref
Murakami, T. {\it et al.} 1994, Nature 368, 127
\ref
Paczy\'nski, B. 1986, Ap. J. 308, L43
\ref
Paczy\'nski, B. 1992, Acta Astr. 42, 145
\ref
Rothschild, R. E., Kulkarni, S. R. and Lingenfelter, R. E. 1994, Nature 368,
432
\ref
Smith, I. A. and Epstein, R. I. 1993, Ap. J. 410, 315
\ref
Sonobe, T. {\it et al.} 1994, Ap. J. 436, L23
\ref
Ulmer, A. {\it et al.} 1993, Ap. J. 418, 395
\vfil
\eject
\end
\bye